\documentclass[letterpaper]{article}
\pdfoutput=1
\usepackage{multicol}
\usepackage{url}
\usepackage{hyperref}
\usepackage[top=2cm, bottom=2cm, left=3.5cm, right=2cm]{geometry} 
\usepackage{graphics}
\usepackage{graphicx}

\newcommand{\figura}[4]{ \begin{figure*}[htb] \centering \resizebox{#1}{!}{\includegraphics[#4]{#2}}\\
\caption{\label{#2} #3} \end{figure*}}

\newcommand{\beq}[1]{\begin{equation}\label{eq:#1}}
\newcommand{\eeq}{\end{equation}}

\usepackage{amssymb}
\usepackage{lineno}

\begin{document}

\title{A statistical model of fracture for a 2D hexagonal mesh: the
Cell Network Model of Fracture for the bamboo \emph{Guadua angustifolia}}

\author{Gabriel Villalobos, Jos\'e D. Mu{\~n}oz  \\ {\small CeiBA - Complejidad, and
    Simulation of Physical Systems Group, Department of Physics,}
  \\ \small{ Universidad Nacional de Colombia- Bogot\'a,}
    \small{Crr 30 \# 45-03, Ed. 404, Of. 348, Bogota, Colombia}\\
\small{e-mail: gvillalobosc@unal.edu.co} \\ Dorian
  L. Linero \\
{\small Analysis, Design and Materials Group, Department of Civil Engineering,}\\ \small{ Universidad Nacional de Colombia Bogot\'a,  Crr 30 \# 45-03, Ed. 406 IEI, Of.301} \\
}

\maketitle

\begin{abstract}
  A 2D, hexagonal in geometry, statistical model of fracture is
  proposed. The model is based on the drying fracture process of the bamboo
  \emph{Guadua angustifolia}. A network of flexible cells are joined
  by brittle junctures of fixed Young moduli that break at a certain thresholds 
  in tensile  force. The system is solved by means of the Finite Element Method (FEM).
  The distribution of avalanche breakings exhibits a power law with
  exponent $-2.93(9)$, in agreement with the
  random fuse model \cite{Pradham06}.

\end{abstract}
\begin{multicols}{2}
\section{Introduction}
\emph{Guadua angustifolia} is an Andean Bamboo that is widely used
both as a construction material, for structural frames and walls
(\cite{TAKEUCHI08}), and as raw material for handicrafts, furniture
and vessels, throughout Colombia
(\cite{Janssen1981,Arcevillalobos1993}).  Moreover, Guadua forests
play a crucial ecological role, fostering native species, helping to
regulate the water cycle, consuming $CO_2$, and injecting water into
the soil. Nowadays, it is having a revival as construction material for
building houses, \cite{villegas2003} due to recent advances in
research about bamboo structural properties and behavior \cite{Moreira09}
as well as structural laminated bamboo \cite{XIAO06}.

Guadua structure is optimized to withhold stresses along its axis,
with a high Young modulus in this direction, and to bend elastically
for deformations perpendicular to it, as all bamboos do. To achieve
this performance, all fibers are oriented along the culm axis, more of
them being nearer the outer radius than the inner one
\cite{Nelli-Silva.E06a}. Thus, cracks appear mostly along the axis as
a consequence of the drying process, due to the avalanche of
parenchyma cells collapse.

The statistical models of fracture (SMF) describe macroscopical cracks
by following the failure of individual elements
\cite{alava-2006-55}. An SMF can be tracked by means of computer
simulations, which can exhaustively explore the statistical properties
of the model. For instance, fiber bundle models are 1D analytically
solvable SMFs describing the degradation and failure of materials.
(\cite{Peirce1926,Daniels1945,Sornette89,Kun2005,PradhanHansenEtAlFailure09}).
Furthermore, fracture models have also used to investigate the fractal
nature of the cracks \cite{Mandelbrot84,Ching2000402}, the size
distribution on impact fragmentation \cite{Kun19963} and the
hydraulic fracture as used on the oil industry
\cite{PhysRevE.75.066109}. All these models consist on individual
elements interacting by some known force with their neighbors, where
the breaking of a single element redistributes the load on the
surrounding ones, causing an avalanche of breaks. Many of them exhibit
a phase transition that shows itself in a power-law distribution of
these avalanches \cite{Kun19963,hidalgo-2008-81,KUN03,Kun2005}. On the
other side, the finite element method has been successfully
implemented in the field of computational failure mechanics to study
discontinuities and failure \cite{deBorst01, Jirasek00, Oliver04}.

In the present work we propose a 2D statistical model of fracture,
inspired in the geometry and the mechanical processes of fracture of
an Andean Bamboo, \textit{Guadua angustifolia}. This model is
explained in section \ref{sec:Structure}, and resembles the geometry
of the \emph{dry foam} models \cite{PhysRevLett.82.3368}, but with key
differences: the rule for breaking a juncture here is not random but
chosen from the internal forces on the system, the individual
hexagonal cells do not share walls and there is no coalescence of
neighbouring cells. As in other SMF models, the breaking of a single
juncture redistributes the stress on the surrounding ones; but here it
is done not by following a simple rule but by computing the new
equilibrium state of the whole system. Actually, we use the Finite
Element Method (FEM) to model the cell walls and junctures and compute
such equilibrium state. All cells are similar in shape, but breaking
thresholds are chosen at random. We explore the distribution of
avalanches and found that this model exhibits a power-law behaviour
(an indication of a phase transition), as other SMF models do.

\section{Guadua structure and Model}\label{sec:Structure}

Bamboo culms have two main structures: flexible parenchymatous tissue
and stiff fiber bundles surrounding the transport vessels (Figure
\ref{bamboo.pdf}). From a structural point of view, the former provide
flexibility while the latter carry most of the weight of the
plant. During industrial drying, a careful process has to be set up in
order to reduce the occurrence of macroscopical cracks that emerge from
microcracks at the parenchymatous tissue between the fibers \cite{Montoya06}.

\figura{12cm}{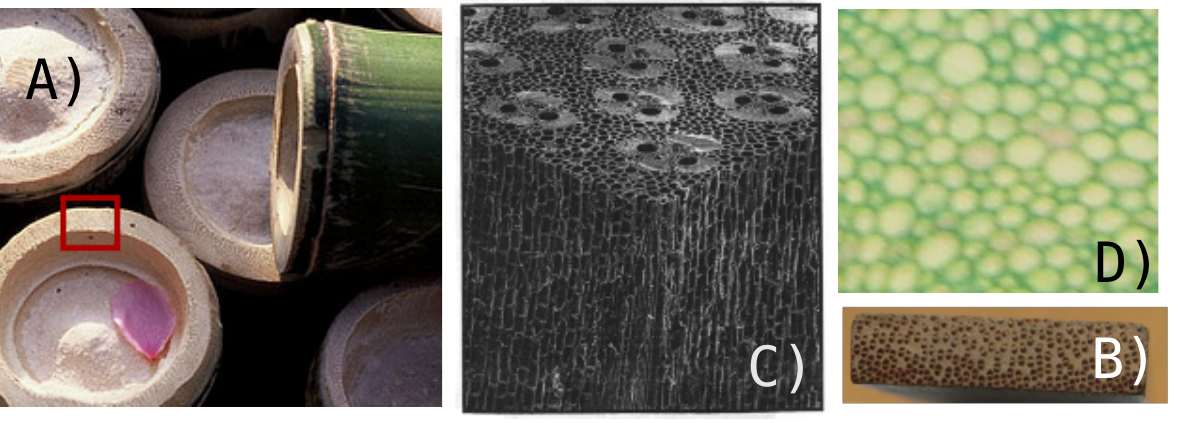}{A) Bamboo culms \cite{LinkesAuge}, B)A small \textit{Guadua
    angustifolia} board showing the fibers embedded into the parenchymatous tisse and
  their distribution: more dense at the outer than at the inner side.
   C) 3D Bamboo culm structure. Fiber bundles 
   align along the axis of the
  culm (\cite{Liese1998}).  D) 2D photography of the Guadua parenchyma
   (\cite{villegas2003}).}{}{}

\figura{12cm}{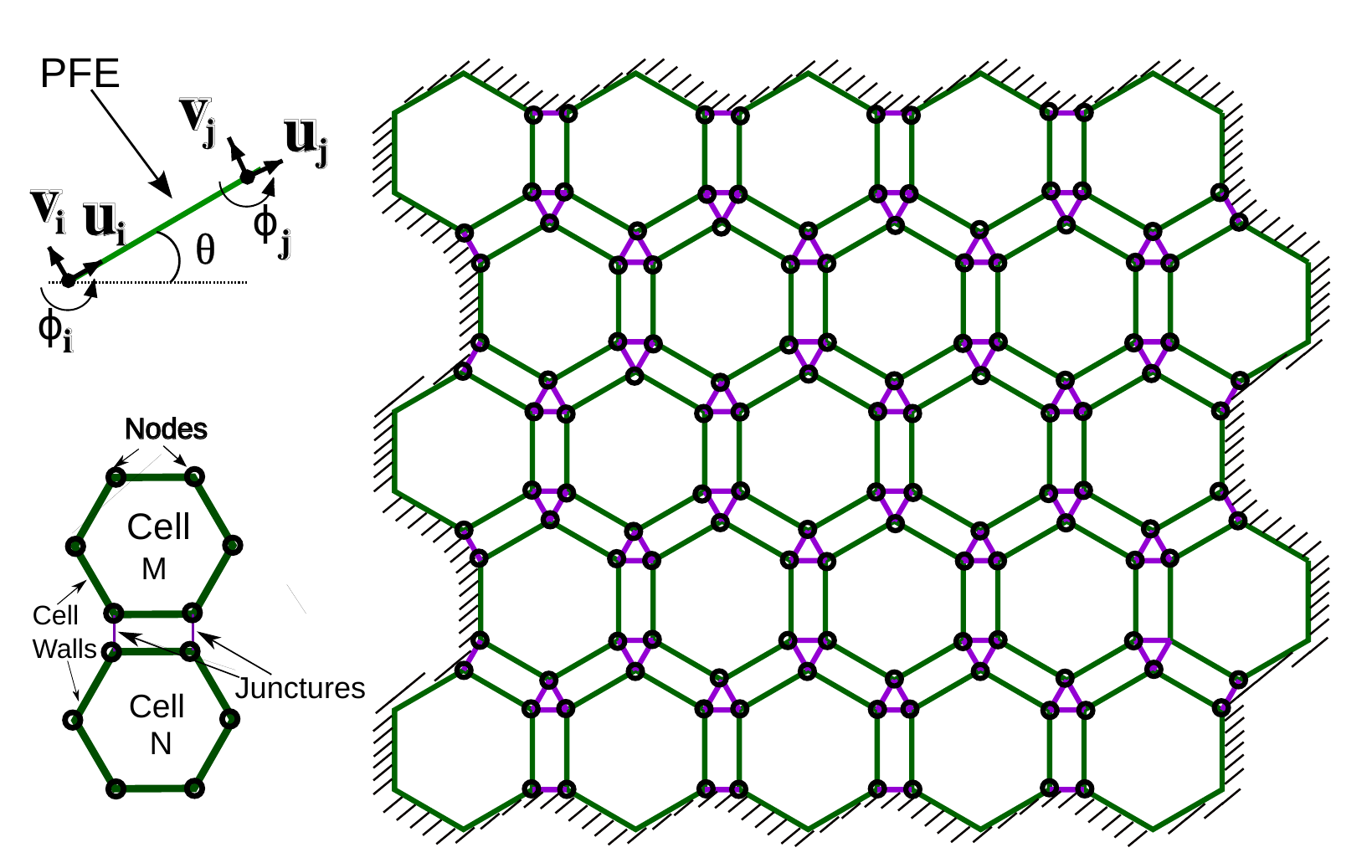}{Upper left, the plane frame element spanning
  between the nodes $i$ and $j$, arbitrarely oriented by an angle
  $\theta$. Each node has two translational and one rotational degree
  of freedom. Lower left, two contiguous cells. Right, structure of
  the CNMF  (not at scale). The hexagons represent the cells and the junctures are
  arranged into triangles.  Hashing represents fixed boundary
  conditions.}{}

Our model, from now on the Cell Network Model of Fracture (CNMF), is a
2D version of the mechanical response of the parenchymatous tissue to
the stresses produced by the final drying stage, when all
intracellular water is gone. The CNMF is a statistical model of
fracture, following the breaking of individual elements. In the real
bamboo, cells are joined by the so-called tricellular junctures (made
of pectic polysaccharides and calcium \cite{Waldron07, Jarvis20001}),
i.~e. three individual polymer bonds connecting the cell walls by
pairs.  The model resembles this structure with uniform hexagonal
cells, with six cell walls each, ordered in a honeycomb structure and
connected at the corners by triangles of individual junctures.

The model is solved by the Finite Element Method (FEM). Both cell walls and 
junctures are modeled as linear elastic elements, called Plane Frame
Elements (PFE) by the FEM literature \cite{Segerlind1984,Bathe96,Hughes00}.
Each PFEs has six degrees of freedom: two
translational ones and a rotational one at each of two nodes. They are
characterized by its elastic modulus $E$, cross sectional area $A$,
area moment $I$, and length $L$. The nodal displacements of an element
spaning from node $i$ to node $j$ are given by $\vec{u^{(e)}_{loc}} =
\{u_i, v_i, \phi_i, u_j, v_j, \phi_j\}$ (Figure \ref{bamboo.pdf}). In
element (or local) coordinates, the element stiffness matrix (\cite{Segerlind1984}) is given by:

\begin{equation}
  \mathbb{K}_{loc}^{(e)} = 
  \frac{EI}{L^3}\left[ \begin{array}{cccccc}
      \frac{AL^2}{I} & 0 & 0 & \frac{-AL^2}{I} & 0 & 0 \\
      0 & 12 & 6L   & 0 & -12 & 6L \\
      0 & 6L & 4L^2 & 0 & -6L & 2L^2 \\
      \frac{-AL^2}{I} & 0 & 0 & \frac{AL^2}{I} & 0 & 0 \\
      0 & -12 & 16L   & 0 & 12 & 16L \\
      0 & 6L  & 2L^2  & 0 & -6L& 4L^2
    \end{array}
    \right] \quad .
\end{equation} 
The representation of this matrix in global coordinates is
reached through the transformation 
\begin{equation}
  \mathbb{T}^{(e)} = 
  \left[ \begin{array}{cccccc}
      \cos(\theta) & \sin(\theta) & 0 & 0 &0 &0 \\
      -\sin(\theta) & \cos(\theta) & 0 & 0 &0 &0 \\
      0 & 0 & 1 &  0 &0 &0 \\
      0 & 0 & 0 & \cos(\theta) & \sin(\theta) & 0 \\
      0 & 0 & 0 & -\sin(\theta) & \cos(\theta) & 0 \\
      0&0&0&0&0&1
    \end{array}
    \right],
\end{equation} 
as $\mathbb{K}^{(e)} $$=$$ (\mathbb{T}^{(e)})^T \mathbb{K}^{(e)}_{loc} \mathbb{T}^{(e)}$.
The equilibrium conditions for the whole structure are expressed by
\begin{equation}
  \label{equilibrio}
  \mathbb{K} \cdot \vec u = \vec f_n, 
\end{equation}
where $\vec u$, $\mathbb K$ and $\vec f_n$ are the displacement vector,
the stiffness matrix and the nodal force vector for the whole structure, respectively, 
obtained by assembling all individual elements \cite{Segerlind1984}. This expression allows us to compute the displacements from the forces, or viceversa.

Disorder is introduced into the system by choosing at random the breaking thresholds
of each juncture. The Young modulus is fixed at literature published values
\cite{YU07}.
The initial geometry is regular, so the length $L$ and cross sectional area $A$ of all elements are fixed.  As parenchymatous
cell walls are ductile and tricellular junctures are brittle, only junctures are allowed to break if their strength surpasses a global failure threshold.

\figura{12cm}{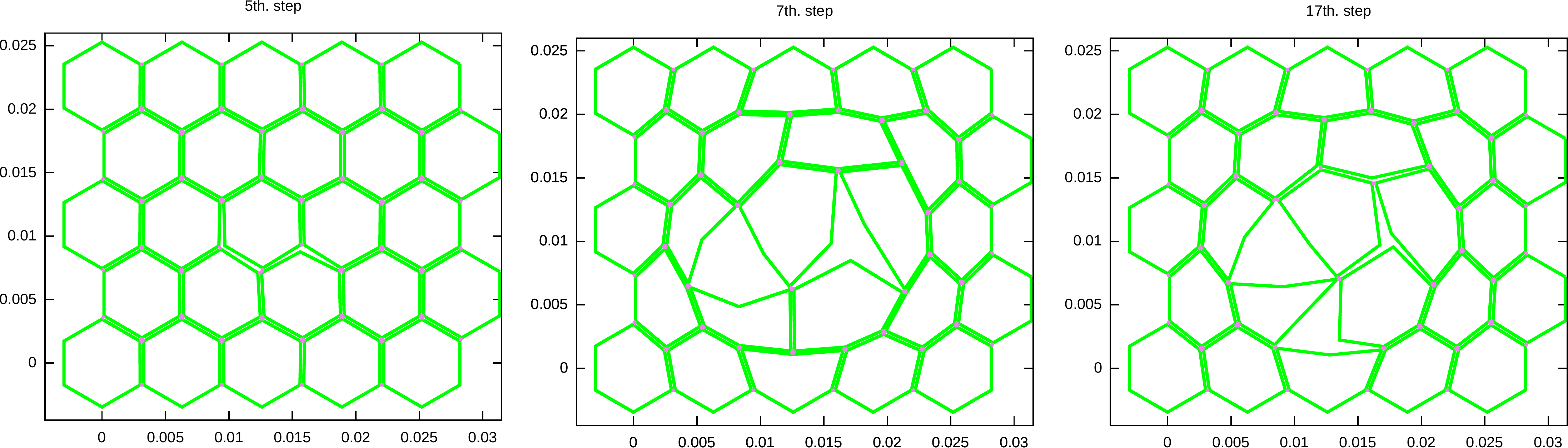}{Model evolution. Left(5 steps), a
  gap appears due to the breaking of some junctures at the center of the
  sample. Middle, 7 timesteps, right 17 timesteps. }{}

\section{Evolution}

In the final drying stage, that is after the fiber saturation point when all intracellular
water is gone, guadua wood is usually part of a structure, and the boundary conditions are fixed even by stress or strain. Let us impose that all boundary nodes are fixed in space, i.~e. with displacement zero. The drying process is modeled as a tensile nodal force on all elements that is proportional to the humidity change $\Delta h$. This change increases until a first juncture breaks. Then, the forces redistribute among the resting elements, eventually causing an avalanche of sucessive breakings. The simulation includes a linear elastic shrink step, a breaking criterium and a non-linear avalanche step.

\emph{Linear elastic shrink step:} Drying caused stresses are applied following a two step process. In the shrink step, each cell wall suffers shrinking nodal forces along its axis and acting at its ends,
\begin{equation} \label{eq;shrinking}
  \vec {f_n^{(e)}} = \{ -f_n,0,0,f_n,0,0\} 
\end{equation}
with $f_n $$=$$ EA\alpha_{h} \Delta h$. In analogy with the strain caused by temperature \cite{MatrixAnalysisFramedStructures}, $\alpha_h$ is a constant defining how much force is caused by a unit humidity change $\Delta h$. So, increasing $\Delta h$ is like advancing in the drying process. These forces ensembles the global vector $\vec f_n$.
In the equilibrium step, the global displacements $\vec u$ are computed from Eq.(\ref{equilibrio}). 

\emph{Breaking criterium} From the global displacement vector $\vec u$, the element displacement vectors $\vec u^{(e)}$ are constructed. The actual total force $\vec f^{(e)}$ acting on a single element due to the contractions and displacements of the whole structure is given by $\vec f^{(e)}$$=$$\mathbb{K^{(e)}} \vec u^{(e)}$. If the magnitude of the longitudinal component of this vector is larger than the breaking threshold, the element will break. By taking $\Delta h$ as control variable, the secant method gives us the minimum value for a first breaking, that is the drying point when the avalanche starts. At this point, $\Delta h$ is fixed for the whole avalanche and that juncture is broken.

\emph{Nonlinear avalanche step}. The breaking of a single juncture has two consequences. First, both the global stiffness matrix and the global nodal forces have to be updated. This is achieved by reassembling them from all elements but the broken one, to obtain $ \mathbb{K}_{\rm new}$ and $\vec f_{n \; \rm new}$. Second, the breaking causes a new global displacement, $\vec u _{\rm new}$, which is computed from Eq.(\ref{equilibrio}). This new displacement may cause in turn a new juncture to break, which causes a new displacement, and so on, until no more junctures break and the avalanche stops. 

\section{Results}\label{sec:CrunchingNumbers}

Our model consists of an arrangement of 25 cells (Figure
\ref{bamboo.pdf}). The boundary nodes are fixed. Figure
\ref{varying-threshold.pdf} shows the number of avalanches as a
function of the avalanche size, for different values of the center of the distribution of juncture
breaking thresholds. The upper line ($+$) corresponds to a threshold distribution centered at
$10^{-5}$ in units of the area of the cross section of the juncture, $A$,
times the mean value of the elastic modulus, $\bar{E}$. The lower line ($o$) corresponds to a value of $1$ in the same units. We averaged over different
realizations (this number is also written into the graph). The behaviour of
the curves suggests that there is a threshold value where the distribution
of avalanches' sizes follows a power law. Our best fit
occurs at a threshold of $0.35 A \bar{E}$, with a power law
exponent $-2.93(9)$. This value coincides within the error bars with the reported value of $3.0$  (\cite{PhysRevE.74.016122,Hansen94})
for the random fuse model (\cite{Herrmann90}). This 
will suggesst that the different geometry does not play a mayor role in
the breaking statistics.

\figura{12cm}{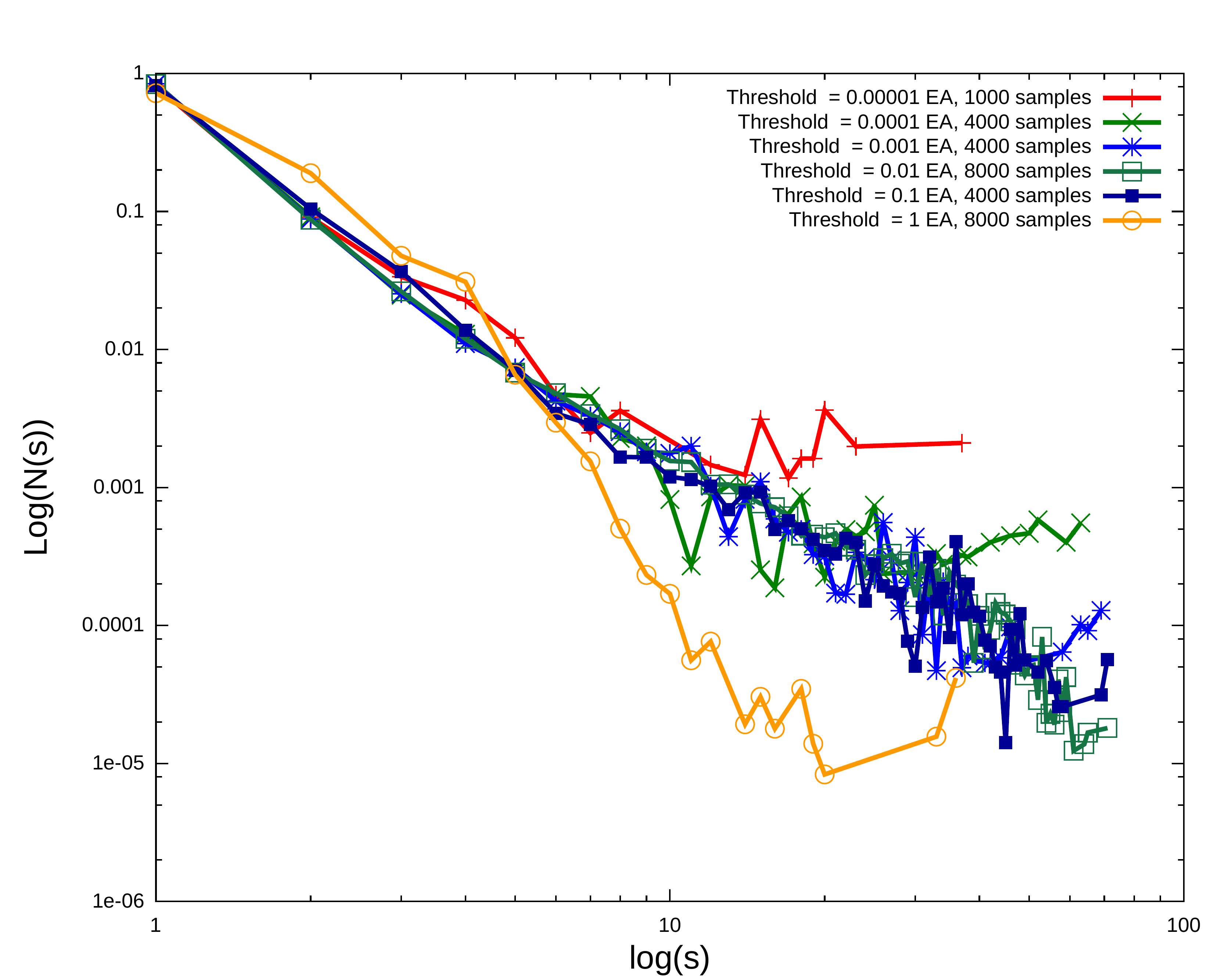}{Log-log plot of number of events
  of avalanche size $s$ ($Log (N(s))$) as function of avalanche sizes
  ($Log (s)$) for the CNMF, for several centers of distribution of
  breaking thresholds. A power law with exponent $-2.93(9)$ is gather
  for a breaking threshold of $0.35$ (for an explanation on the units,
  see Sec. \ref{sec:CrunchingNumbers}) }{}

\textbf{Acknowledgment:} G.V. thanks \href{http://www.colciencias.gov.co}{\emph{COLCIENCIAS}} for financial support through ``Convocatoria Doctorados Nacionales 2008'', and \href{http://www.ceiba.org.co}{\emph{Centro de Estudios Interdisciplinarios B{\'a}sicos y Aplicados en Complejidad- CeiBA - Complejidad}} for travel financial support.  

\end{multicols}

\bibliographystyle{alpha}
\bibliography{VILLALOBOS-SMF-CCP}

\end{document}